\documentclass[osajnl,preprint]{revtex4}

\usepackage{graphicx}
\usepackage{amsmath}
\usepackage{bm}
\usepackage{mathptmx}

\begin{document}

\title{Filamentational Instability of Partially Coherent Femtosecond Optical Pulses in Air}

\author{M. Marklund and P. K. Shukla}
\affiliation{Centre for Nonlinear Physics, Department of Physics, 
Ume{\aa} University, SE--901 87 Ume{\aa}, Sweden}
\affiliation{Institut f\"ur Theoretische Physik IV and Centre for Plasma Science and Astrophysics,
Ruhr-Universit\"at Bochum, D-44780 Bochum, Germany}

\date{Revised 21 March 2006, accepted for publication in Opt.\ Lett.}

\begin{abstract}
The filamentational instability of spatially broadband femtosecond optical pulses 
in air is investigated by means of a kinetic wave equation for spatially incoherent photons. 
An explicit expression for the spatial amplification rate is derived and analyzed.
It is found that the spatial spectral broadening of the pulse can lead to 
stabilization of the filamentation instability. Thus, optical smoothing techniques 
could optimize current applications of ultra-short laser pulses, such as 
atmospheric remote sensing.
\end{abstract}

\maketitle

Recently, there has been a great deal of interest \cite{r1,r2,r3,r4,r5,r6,r7,r8} in investigating 
the nonlinear propagation of optical pulses in air. In order for the pulse propagation over
a long distance, it is necessary to avoid filamentational instabilities that grow in space.
Filamentation instabilities of optical pulses occur in nonlinear dispersive media, 
where the medium index of refraction depends on the pulse intensity. This happens 
in nonlinear optics (viz. a nonlinear Kerr medium) where a small modulation of the 
optical pulse amplitudes can grow in space due to the filamentation instability arising 
from the interplay between the medium nonlinearity and the pulse dispersion/diffraction. 
The filamentational instability is responsible for the break up of pulses into light pipes. 
It is, therefore, quite important to look for mechanisms that contribute to the nonlinear 
stability of optical pulses in nonlinear dispersive media. One possibility would be to 
use optical pulses that have finite spectral bandwidth, since the latter can significantly 
reduce the growth rate of the filamentation instability. Physically, this happens because
of the distribution of the optical pulse intensity over a broad spectrum, which is unable
to drive the filamentation instability with fuller efficiency, contrary to 
a coherent pulse which has a delta-function spectrum. In this Letter, we present 
for the first time a theoretical study of the filamentation instability of partially 
coherent optical pulses in air.  We show that the spatial amplification rate of 
the filamentation instability is significantly reduced by means of spatial spectral broadening 
of optical pulses. The present results could be of significance in applications 
using ultra-short pulses for remote sensing of the near Earth atmosphere.   

The dynamics of coherent femtosecond optical pulses with a weak group velocity dispersion 
in air is governed by the modified nonlinear Schr\"odinger equation 
\cite{r4,r5,Berge-etal,Vincotte-Berge,Skupin-etal}
\begin{equation}\label{eq:nlse}
  i\partial_z\psi +\nabla_{\perp}^2\psi + f(|\psi|^2)\psi + i\nu|\psi|^{2K - 2}\psi = 0 ,
\end{equation}
where $\psi(z,\mathbf{r}_{\perp})$ is the spatial wave envelope, $\mathbf{r}_{\perp} = (x,y)$, 
and $f(|\psi^2|) = \alpha|\psi|^2 - \epsilon|\psi|^4 - \gamma|\psi|^{2K}$. Here 
$\alpha = 0.466$, $\epsilon = 7.3\times 10^{-7} \, \mathrm{cm}^2/w_0^2$,
$\gamma = 8.4\times 10^{-40} \,\mathrm{cm}^{2(K - 1)}/w_0^{2(K - 1)}$, and 
$\nu = 1.2\times 10^{-35}\,\mathrm{cm}^{2(K - 2)}/w_0^{2(K - 2)}$ for a
pulse duration of $250\,\mathrm{fs}$, and $w_0$ (in units of $\mathrm{cm}$) is the 
beam waist  \cite{Vincotte-Berge} (for a discussion of the approximations leading 
to Eq.\ (\ref{eq:nlse}), we refer to \onlinecite{Berge-etal}).  We note that Eq. (\ref{eq:nlse}) 
has been used in Ref. \onlinecite{Skupin-etal} to analyze the multi-filamentation of optical beams. 

Following Ref. \onlinecite{Fedele}, we can derive a wave kinetic equation that governs the nonlinear 
propagation intense optical pulses which have a spectral broadening in space. Accordingly, we apply 
the Wigner-Moyal transform method \cite{Wigner,Moyal,Ivleva-etal,Korobkin-Sazanov}. The multi-dimensional
Wigner-Moyal transform, including the Klimontovich statistical average, is defined as
\begin{equation}\label{eq:wignerfunc}
  \rho(z,\mathbf{r}_{\perp},\mathbf{p}) = 
  \frac{1}{(2\pi)^2}\int\,d^2\xi\,e^{i\mathbf{p}\cdot\bm{\xi}}\langle  
    \psi^*(z,\mathbf{r}_{\perp} + \bm{\xi}/2)\psi(z,\mathbf{r}_{\perp} - \bm{\xi}/2)
  \rangle, 
\end{equation}
where $\mathbf{p} = (p_x,p_y)$ represents the momenta of the quasiparticles and the angular bracket 
denotes the ensemble average \cite{Klimontovich}. The pulse intensity $\langle|\psi|^2\rangle \equiv I$ 
satisfies
\begin{equation}\label{eq:intensity}
  I = \int\,d^2\!p\,\rho(z,\mathbf{r}_{\perp},\mathbf{p}) .
\end{equation} 
Applying the transformation (2) on Eq. (\ref{eq:wignerfunc}), we obtain the Wigner-Moyal 
kinetic equation \cite{Moyal,Ivleva-etal,Korobkin-Sazanov,Mendonca} for the evolution of the 
Wigner distribution function,
\begin{equation}\label{eq:wigner}
  \partial_z\rho + 2\mathbf{p}\cdot\nabla_{\perp}\rho 
  + 2f(I)\sin\left( \tfrac{1}{2}\stackrel{\leftarrow}{\nabla}_{\perp}\cdot\stackrel{\rightarrow}
  {\nabla}_{p}\right)\rho
  + 2\nu I^{K - 1}\cos\left( \tfrac{1}{2}\stackrel{\leftarrow}{\nabla}_{\perp}\cdot\stackrel
  {\rightarrow}{\nabla}_{p}\right)\rho = 0 .
\end{equation}

Seeking the solution $\bar{\rho} = \bar{\rho}(z,\mathbf{p})$ to Eq.\ (\ref{eq:wigner}), we may write 
$\bar{\rho}(z,\mathbf{p}) = \rho_0(\mathbf{p})\bar{I}(z)$, where $\rho_0$ is an arbitrary function 
of $\mathbf{p}$ satisfying $\int\,d^2p\,\rho_0 = 1$, and $\bar{I}(z) 
= I_0(2K - 2)/[2\nu I_0^{2K - 2}z + (2K - 2)^{2K - 2}]^{1/(2K - 2)}$, with $I_0 = \bar{I}(0)$. 
Thus, the effect of a small but non-zero $\nu$ is to introduce a slow fall-off in the intensity along 
the $z$-direction when $K \geq 1$. Moreover, as $\nu \rightarrow 0$ this solution reduces to $\bar{I} = I_0$.

We now consider spatial filamentation of a well defined optical pulses against small perturbations 
having the parallel wavenumber $k_{\|}$ and the perpendicular wavevector $\mathbf{k}_{\perp}$, 
by assuming that $\nu$ is small so that $k_{\|} \gg |\partial_z|$ for the background distribution. 
We let $\rho = \bar{\rho}(z,\mathbf{p}) + \rho_1(\mathbf{p})\exp(ik_{\|}z + i\mathbf{k}_{\perp}
\cdot\mathbf{r}_{\perp}) + \mathrm{c.c.}$\ 
and $I = \bar{I}(z) + I_1\exp(ik_{\|}z + i\mathbf{k}_{\perp}\cdot\mathbf{r}_{\perp}) +\mathrm{c.c.}$, 
where $|\rho_1| \ll \bar{\rho}$, $|I_1| \ll \bar{I}$, and $c.c.$ stands for the complex conjugate. 
We linearize (4) with respect to the perturbation variables and readily obtain the nonlinear 
dispersion equation
\begin{equation}\label{eq:disprelgen}
  1 = \int d^2p\,\frac{[f'(\bar{I}) + i\nu(K - 1)\bar{I}^{K-2}]\bar{\rho}(z,\mathbf{p} - \mathbf{k}_{\perp}/2)
    -[f'(\bar{I}) - i\nu(K - 1)\bar{I}^{K-2}]\bar{\rho}(z,\mathbf{p} + \mathbf{k}_{\perp}/2)}%
      {k_{\|} + 2\mathbf{k}_{\perp}\cdot\mathbf{p} - 2i\nu\bar{I}^{K - 1}} ,
\end{equation}  
which is valid for partially coherent femtosecond pulses in air. Here the prime denotes differentiation with 
respect to the background intensity $\bar{I}$.

We simplify the analysis by assuming that the perpendicular dependence in essence is one-dimensional. 
In the coherent case, i.e.\ $\bar{\rho}(z,p) = \bar{I}(z)\delta(p - p_0)$, Eq. (\ref{eq:disprelgen}) 
yields 
\begin{equation}\label{eq:monodisprel}
  k_{\|} = -2kp_0 + i\nu(K + 1)\bar{I}^{K - 1} \pm \sqrt{%
    k^2[k^2 - 2f'(\bar{I})\bar{I}] - \nu^2(K - 1)^2\bar{I}^{2K - 2}},
 \end{equation}
where $k$ represents the perpendicular wavenumber in the one-dimensional case.
Letting $k_{\|} = -2kp_0 - i\Gamma$ in (6), where $\Gamma$ is the filamentation instability growth 
rate, we thus obtain \cite{Couairon-Berge}
\begin{equation}\label{eq:growthrate}
  \Gamma = -\nu(K + 1)\bar{I}^{K - 1} + \sqrt{%
    k^2[2f'(\bar{I})\bar{I} - k^2] + \nu^2(K - 1)^2\bar{I}^{2K - 2}%
  },
\end{equation} 
which reduces to the well known filamentation instability growth rate in a Kerr medium 
(i.e.\ $\nu = 0$ and $f(I) = \alpha I$). We note that a nonzero $\nu$ gives rise to an overall 
reduction of the growth rate. In Fig.\ 1 we have plotted a number of different curves for 
the growth rate in the coherent case.

In the partially coherent case, we investigate the effects of spatial spectral broadening using the 
Lorentz distribution
\begin{equation}\label{eq:lorentz}
  \bar{\rho}(z,p) = \frac{\bar{I}(z)}{\pi} \frac{\Delta}{(p-p_0)^2 + \Delta^2} ,
\end{equation}
where $\Delta$ denotes the width of the distribution around the quasiparticle momenta $p_0$. 
Inserting (\ref{eq:lorentz}) into (\ref{eq:disprelgen}) and carrying out the integration in
a straightforward manner, we obtain
\begin{equation}\label{eq:disprellorentz}
  k_{\|} = -2kp_0 + i\nu(K + 1)\bar{I}^{K - 1} + 2ik\Delta \pm \sqrt{%
    k^2[k^2 - 2f'(\bar{I})\bar{I}] - \nu^2(K - 1)^2\bar{I}^{2K - 2}%
  } . 
\end{equation}
With $k_{\|} = -2kp_0 - i\Gamma$ the filamentation instability growth rate is
\begin{equation}\label{eq:growthrategen}
  \Gamma = - \nu(K + 1)\bar{I}^{K - 1} - 2k\Delta + \sqrt{%
    k^2[2f'(\bar{I})\bar{I} - k^2] + \nu^2(K - 1)^2\bar{I}^{2K - 2} }. 
\end{equation}
In the limit $\Delta \rightarrow 0$, Eq.\ (\ref{eq:disprellorentz}) reduces to the dispersion relation 
(\ref{eq:monodisprel}), while for $\nu = 0$ the dispersion relation (\ref{eq:disprellorentz}) reduces 
to the standard expression for the filamentation instability growth rate 
\begin{equation}
  \Gamma = -2k\Delta + k\sqrt{2I_0f'(I_0) - k^2} .
\end{equation}
In Fig. 2 we have displayed the filamentation instability growth rate (\ref{eq:growthrategen}). 
The effect of the finite width $\Delta$ of the quasiparticle distribution can clearly 
be seen. In particular, multi-photon absorption (here chosen to be a modest 
$K = 3$), determined by the coefficient $\nu$, as well as multi-photon
ionization, represented by the coefficient $\gamma$, combined with finite spectral
width of the optical pulse give rise to a significant reduction of the filamentation
instability growth rate. This is evident from Fig.\ 2, where the plotted normalized growth rate
is reduced by as much as a factor of six, compared to the case of full coherence.

In practice, optical smoothing techniques, such as the use of random phase plates \cite{smooth} 
or other random phase techniques well suited for the results in the present Letter,
have been used in inertial confinement fusion studies for quite some time (see, e.g.\ 
Ref.\ \onlinecite{Koenig-etal}). Such spatial partial coherence controls are reproducible 
and can be tailored as to give a suitable broadband spectrum (as in, e.g. \onlinecite{Moh-etal}, 
where optical vortices were generated).  Thus, in the case of ultra-short pulse propagation in air, 
such random phase techniques can be used to experimentally prepare an ultra-short 
optical pulse for a long-distance propagation, and a large spatial bandwidth of optical 
pulses, in conjunction with multi-photon ionization and absorption, may drastically reduce
(down to less than 20\,\% of the coherent value in the present study) the filamentation 
instability growth rate.  This will lead to a greater long range stability, since the
onset of strong optical pulse filamentation is delayed, resulting in several times longer 
stable propagation.  A rough estimate based on the numbers found in the present Letter shows 
that an optical beam could propagate a distance as much as six times longer with proper 
random phasing.     

To summarize, we have investigated the filamentation instability of partially coherent 
femtosecond optical pulses in air. For this purpose, we introduced the Wigner-Moyal representation 
on the modified nonlinear Schr\"odinger equation and obtained a kinetic wave equation for
optical pulses that have a spectral bandwidth in wavevector space. A perturbation analysis 
of the  kinetic wave equation gives a nonlinear dispersion relation, which describes 
the filamentation instability (spatial amplification) of broadband optical pulses. 
Our results reveal that the latter would not be subjected to filamentation due 
to spectral pulse broadening. Hence, using partial spatial coherence effects for 
controlling the filamentational instability, femtosecond optical pulse propagation
in air can be improved significantly. The result presented here is also indicative that
optical smoothing techniques, as used in inertial confinement studies, could be 
very useful for ultra-short pulse propagation in air. This can help to optimize current 
applications of ultra-short laser pulses for atmospheric remote sensing over a long 
distance.

\acknowledgments
The authors thank one of the referees for helpful suggestions and comments on a 
previous version, as well as providing valuable references. This research was partially 
supported by the Swedish Research Council.


\newpage

  \noindent 
  Fig.\ 1. {The coherent filamentation instability growth rate, given by (\ref{eq:growthrate}), 
  plotted for different parameter values; all curves with $I_0 = 0.5$, $\alpha = 1$, and $K = 3$. 
  The full thick line represents the standard filamentation instability growth rate for a 
  nonlinear Schr\"odinger equation, i.e.\ $\nu = \epsilon = \gamma = 0$; the thin dashed curve 
  has $\nu = \gamma = 0$, while $\epsilon = 0.5$; the thin dotted curve has $\nu = \epsilon = 0$ 
  and $\gamma = 0.5$; the thin dashed--dotted curve has $\nu = 0$ and $\epsilon = \gamma = 0.5$; 
  the thick dashed curve has $\nu = 0.1$ and $\epsilon = \gamma = 0$; finally, the thick
  dashed--dotted curve has $\nu = 0.1$ and $\epsilon = \gamma = 1/2$.}

\ \\[5mm]

  \noindent
  Fig.\ 2. {The partially coherent filamentation instability growth rate, given by 
  (\ref{eq:growthrategen}), plotted for different parameter values; all curves with 
  $I_0 = 0.75$, $\alpha = 1$, and $K = 3$. The full thick line again represents the 
  standard filamentation instability growth rate for a nonlinear Schr\"odinger equation, 
  i.e.\ $\Delta = \nu = \epsilon = \gamma = 0$; the thin full curve has 
  $\nu = \epsilon = \gamma = 0$, while $\Delta = 0.1$; the thin dashed curve 
  has $\epsilon = \gamma =0$ while $\nu = 0.05$ and $\Delta = 0.1$; the thin dotted curve 
  has $\nu = 0.05$ and $\gamma = 0.1$ while $\epsilon = 0$. The effects 
  finite width of the background intensity distribution of the optical 
  pulse, as well as the influence of the higher order nonlinearity 
  and losses are clearly seen here.}

\newpage

\begin{figure}
\includegraphics[width=0.48\textwidth]{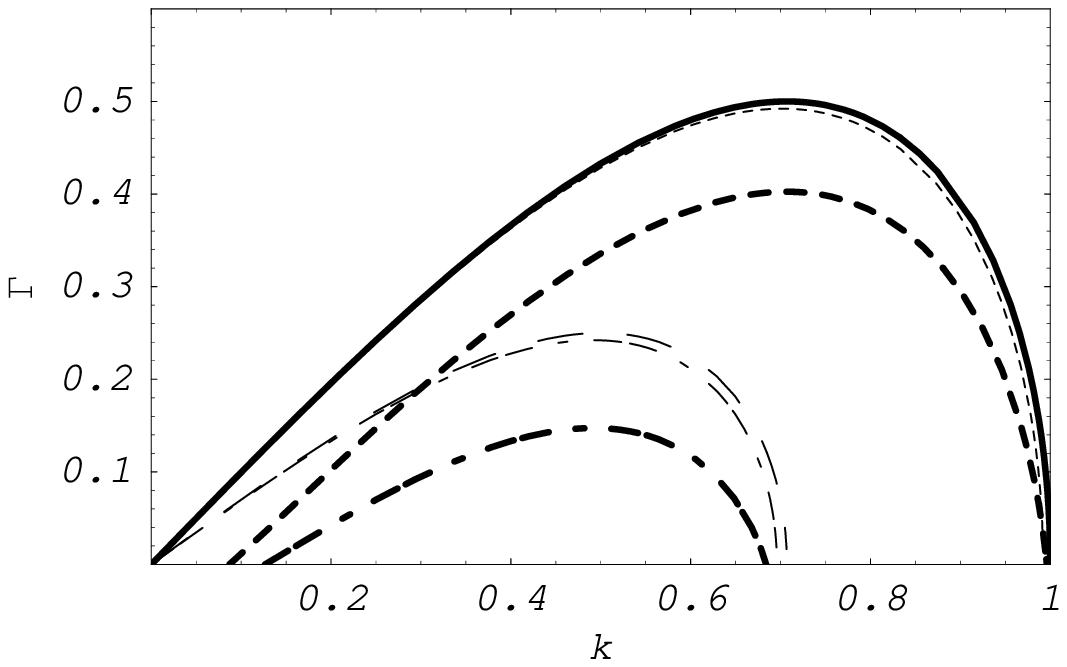}
\caption{}
\end{figure}

\begin{figure}
\includegraphics[width=0.48\textwidth]{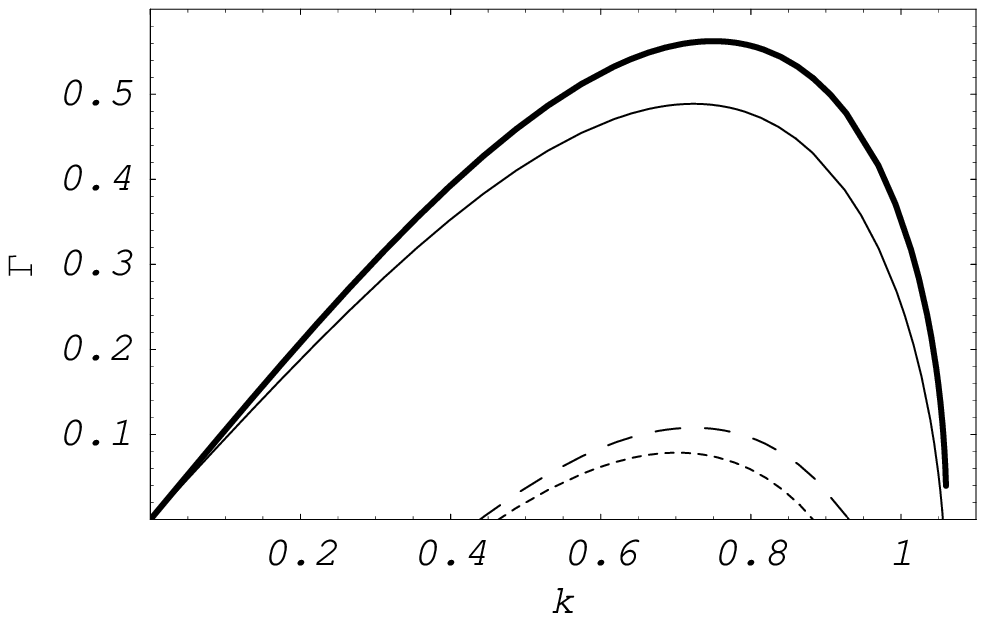}
\caption{}
\end{figure}

\end{document}